\documentclass[12pt]{iopart}

\usepackage{natbib}
\usepackage{amsfonts,amssymb}
\usepackage{epsfig}
\usepackage{xspace}

\def\hypgeo2F2{\mathop{_2{\rm F}_2}}
\def\expint{\mathop{{\rm Ei}}}

\def\rs{r_{\rm s}}
\def\rvir{r_{200}}
\def\rt{r_{\rm t}}

\def\deltac{\delta_{\rm c}}
\def\rhoc{\rho_{\rm crit}}

\def\trunc{\tau}
\def\Li{\mathop{\rm Li}}

\def\reff@jnl#1{{\it#1\/}}
\def\aj{\reff@jnl{Astron. J.}}                  
\def\apj{\reff@jnl{Astrophys. J.}}                
\def\apjl{\reff@jnl{Astrophys. J.}}               
\def\aap{\reff@jnl{Astron. Astrophys.}}               
\def\mnras{\reff@jnl{Mon. Not. Roy. Astro. Soc.}}            
\def\prd{\reff@jnl{Phys. Rev. D}}         
\def\prl{\reff@jnl{Phys. Rev. Lett}}      
\def\pasj{\reff@jnl{ Publ. Astron. Soc. Japan}}              
\def\nat{\reff@jnl{Nature}}             

\begin{document}

\title[NFW lenses]%
{Analytic models of plausible gravitational lens potentials}
\label{firstpage}
\author{
Edward A. Baltz$^1$,
Phil Marshall$^{1,2}$, Masamune Oguri$^1$}
\address{
$^1$ Kavli Institute for Particle Astrophysics and Cosmology, Stanford
  University PO Box 20450, MS29, Stanford, CA 94309, USA \\
$^2$ Physics Department, University of California, Santa Barbara, CA
  93601, USA} 
\ead{eabaltz@slac.stanford.edu}

\begin{abstract}

Gravitational lenses on galaxy scales are plausibly modelled as
having ellipsoidal symmetry and a universal dark matter density profile,  with a
Sersic profile to describe the distribution of baryonic matter. Predicting all
lensing effects requires knowledge of the total lens potential: in this work we
give analytic forms for that of the above hybrid model.
Emphasising that complex lens potentials can be constructed from simpler
components in linear combination, we provide a recipe for attaining elliptical
symmetry in either projected mass or lens potential. We also provide analytic
formulae for the lens potentials of Sersic profiles for integer and
half-integer index.
We then present formulae describing the gravitational lensing effects due to
smoothly-truncated universal density profiles in cold dark matter model. 
For our isolated haloes the density profile falls off as radius to the
minus fifth or seventh power beyond the tidal radius, functional forms
that allow all orders of lens potential derivatives to be calculated
analytically, while ensuring a non-divergent total mass. We show how the
observables predicted by this profile differ from that of the original
infinite-mass NFW profile. Expressions for the gravitational flexion are
highlighted. We show how decreasing the tidal radius allows stripped
haloes to be modelled, providing a framework for a fuller investigation
of dark matter substructure in galaxies and clusters. Finally we remark
on the  need for finite mass halo profiles when doing cosmological
ray-tracing simulations, and the need for readily-calculable higher
order derivatives of the lens potential when studying catastrophes in
strong lenses. 

\end{abstract}


\maketitle


\section{Introduction}
\label{sect:intro}

A large number of cold dark matter (CDM) N-body simulations agree that the
haloes formed have, on average, a universal broken power law density
profile. While there is some debate over the logarithmic slope of the profile
within the break, or scale radius $\rs$, and there is some scatter between
haloes, it seems that the original profile of \cite{NFW97} (NFW) still
provides a reasonable fit to the simulation data. This profile seems to
be a generic feature of haloes formed in the hierarchical model of
structure formation.  

A number of authors have made significant progress in understanding non-linear
effects in structure formation using halo models, where the number density,
correlation function and mass density profile of CDM haloes are fitted to the
numerical simulations and then used in a simplified model of large scale
structure \cite{M+W96,CHM00,T+J03}. The NFW profile has played a
prominent role in this enterprise. One application of the halo model is in the
investigation of the halo occupation distribution, that characterises the
substructure within a larger halo. This has long been one of the more
controversial topics in CDM theory, with predictions and observations often at
odds.  In order to build up an accurate picture of a hierarchical mass
distribution, the stripping of the sub-haloes by tidal gravitational
forces must be modelled \cite{T+B04,O+L04}; measurements of halo
stripping form an important test of the detailed predictions of the
CDM simulations \cite{Hay++03,T+B05}. 

Moreover, it is now clear that the effect of baryons on the shapes and profiles
of total mass distributions cannot be ignored. In galaxies, the stellar
component of the mass dominates at small radii giving rise to a peakier
observed total density profile than seen in pure dark matter simulations
\cite{Koo++06}. The surface brightness profiles of massive galaxies seem to
be consistently well-fitted by a Sersic profile of index $3-4$
\cite{Tru++04}; a logarithmic slope of $-2$ in total density
in the inner regions appears ubiquitous \cite{Tre++06}. Moreover, the
dark matter profile itself is expected to steepen during the formation
of the galaxy, by the process of adiabatic contraction
\cite{Zel++80,Blu++86}. Typically this leads to more centrally
concentrated, rounder haloes \cite{Gne++04,Kaz++04}. The details of the
mass distributions of real galaxies are therefore a probe of the galaxy
formation physics claimed by the simulations. 

Gravitational lensing allows us to probe the mass distributions of galaxies,
groups and clusters in a unique way. Insensitive to the dynamical state of the
lens system, both weak and strong lensing effects depend only on the projected
(and scaled) gravitational potential. Gravitational lensing has already been
used to investigate the density profile in galaxy clusters 
\cite{Kne++03,Gav++03,San++04,Bro++05}. Substructure studies
have also been undertaken, making use of the galaxy-galaxy lensing
effect in clusters \cite{N+S04}. The galaxy scale halo mass profiles
have also been measured, using both strong lensing \cite{RKK03,D+W05,
Koo++06} and, in a more statistical fashion, galaxy-galaxy weak lensing
\cite{She++04, HYG04}. 

Lensing studies provide direct tests of the CDM simulations, and typically
involve (at some point) fitting the parameters of an NFW-like model to the data.
However, this model is also well-suited to a more general analysis, building up
a data model from a linear combination of NFW-like potentials \cite{Mar06}. 
This approach has applications in substructure characterisation, and
also template-based cluster finding. Characterising stripped substructure
both require an accurate treatment of the outer regions of haloes.  However, in
order to measure accurately density profile slopes and concentrations, the
baryonic mass component must be included.

In the perpetually applicable thin lens approximation it is the projected
Newtonian gravitational potential that gives rise to the gravitational lensing
observables.  Making the simplifying assumption that projected stellar mass
density is proportional to optical surface brightness leads us to seek the
potential that corresponds to the Sersic density profile. Likewise, for the
dark matter component we require a lens potential that corresponds to the
universal profile seen in simulations, but that also includes the effects of
tidal stripping.

Analytic lens potentials are convenient to work with: they, and their
derivatives that are needed for lensing data modelling, can be computed quickly
and accurately; the introduction of ellipticity to the halo can be done very
straightforwardly; more complex potentials can be constructed by simple linear
combination of analytic functions. This last feature allows concave isodensity
contours to be avoided in the case of high ellipticity. It also allows total
density distributions to be constructed from mixtures of dark and luminous
matter.

In this work we present analytic forms for a smoothly truncated universal CDM
gravitational potential, and also for the potentials corresponding to a subset
of the Sersic profiles. If the underlying potential is analytic, so are all the
derivatives needed in gravitational lens studies.  An outline of the paper is
as follows. In Sections~\ref{sect:NFW} and~\ref{sect:sersic} we present our
suggested analytic potential models, for both dark and baryonic matter
components, and outline the derivation of the quantities relevant to
gravitational lensing. We leave the full formulae to an appendix, but in
Section~\ref{sect:obs} we plot the predicted observables, and compare them to
those from an unstripped baryon-free NFW form. In Section~\ref{sect:discuss} we
briefly discuss our results, and point the reader towards some publically
available computer code.


\section{Smoothly truncated dark matter haloes}
\label{sect:NFW}

The NFW profile for the CDM density $\rho$ of a halo is
\begin{equation}
\rho(r)=\frac{4\deltac
  \rhoc}{\left(\frac{r}{\rs}\right)\left(1+\frac{r}{\rs}\right)^2}=
\frac{M_0}{4\pi r(r+\rs)^2}.
\label{eq:originalNFW}
\end{equation}
The characteristic overdensity $\deltac$ is the density at the scale radius
$\rs$, in units of the critical density $\rhoc$.  Alternatively, we can express
this as $\rho(\rs)=\deltac\rhoc=M_0/(16\pi\rs^3)$.  The NFW profile is
analytically integrable along the line of sight; the most frequently used
formulae for the weak lensing shear \cite{W+B00} and strong lensing image
positions \cite{Bar96} were derived assuming the integral to extend over all
space. Given that the NFW profile has divergent total mass, \cite{T+J03}
suggest using a modified form that is sharply truncated at the virial
radius. The projection integral is then more realistic, with only mass
physically associated with a finite-sized halo being modelled.

We might expect real CDM haloes to be truncated due to tidal effects; a
step-function density cutoff may not offer a very physical picture of the edges
of haloes. With the \cite{T+J03} mass distribution, the lensing deflection
angle and shear are tractable (if somewhat less simple), and the actual
potential is worse still, involving (at least) polylogarithms.  Also, the
convergence and shear are not differentiable at the truncation radius.  A
power-law cutoff in the potential is more attractive in this regard. We should
insist that the truncated profile match that of NFW as closely as possible
within the tidal radius, which is introduced as the third parameter of the
profile. This is important for the results from previous work on fitting the
outputs from N-body simulations pertaining not only to the density profile but
also the mass function \cite{Jen++01}.

\begin{figure}
\begin{minipage}[t]{0.48\linewidth}
\centering\epsfig{file=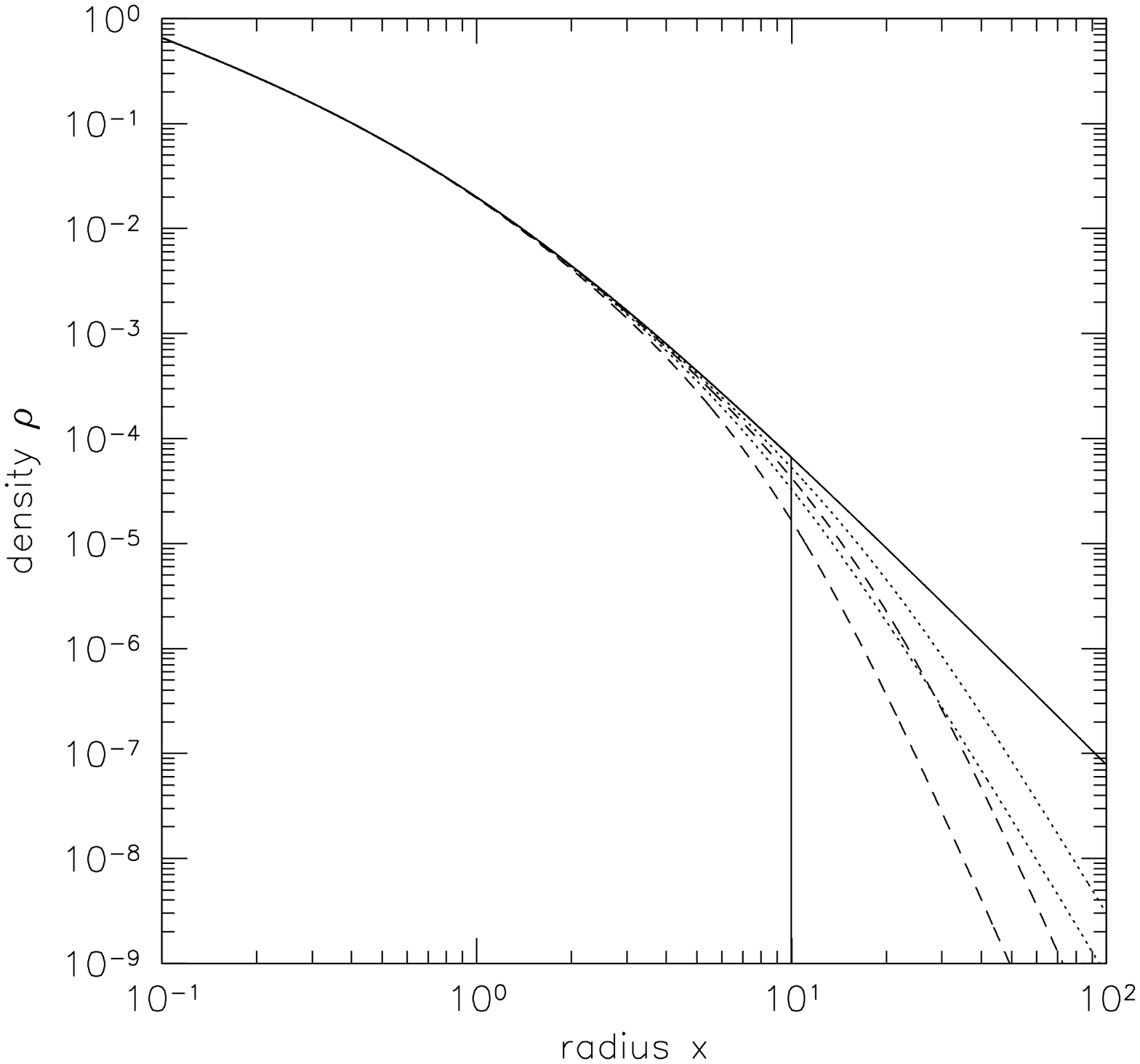,width=0.95\textwidth}
\end{minipage} \hfill
\begin{minipage}[t]{0.48\linewidth}
\centering\epsfig{file=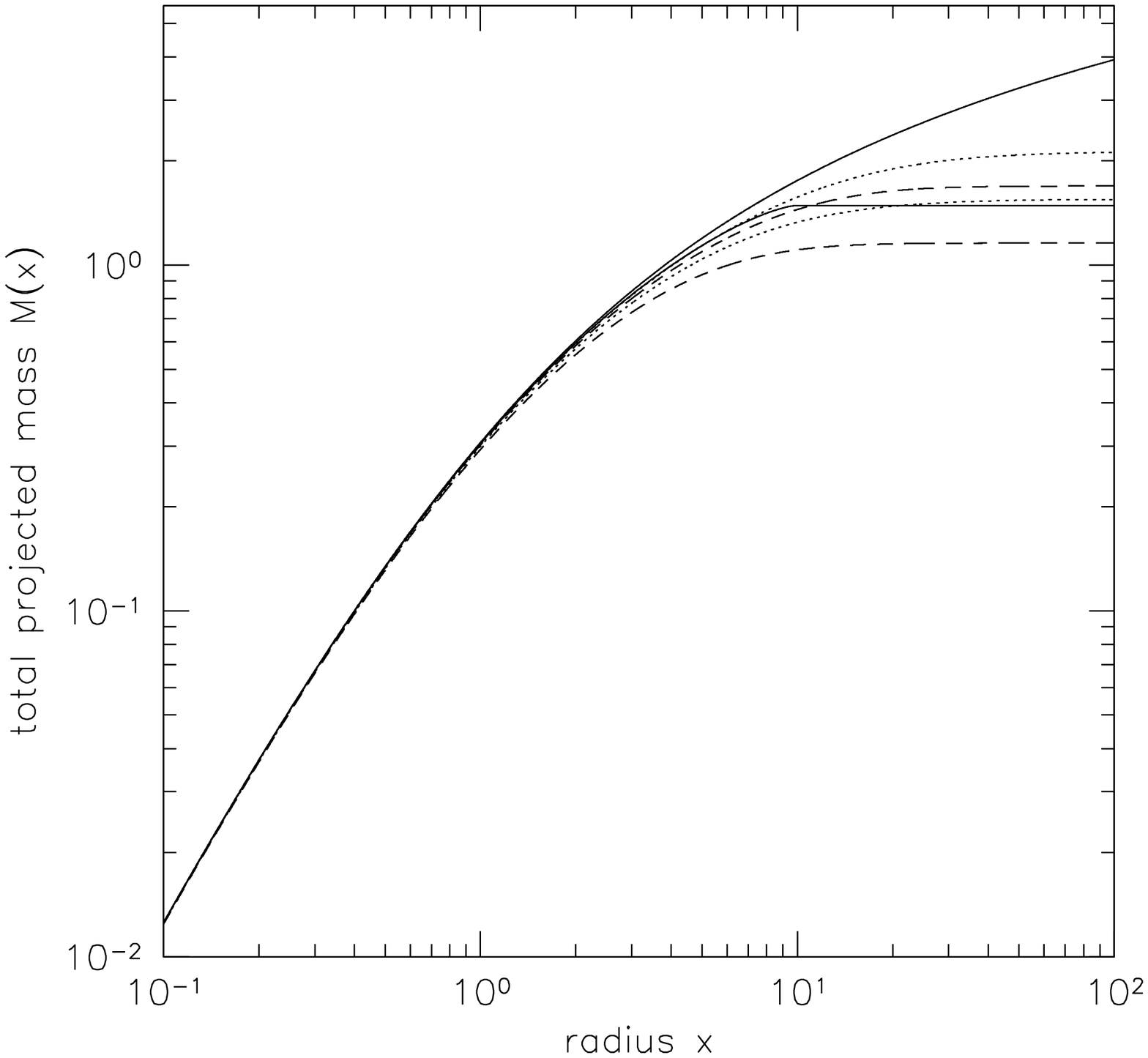,width=0.95\textwidth}
\end{minipage}
\caption{Density (left) and integrated projected mass (right) profiles
  for NFW haloes with various truncation schemes.  The solid lines indicate
  the original NFW halo, with and without a hard cutoff at $\trunc =
  \rt / \rs = 10$.  Dotted lines show the $n=1$ cutoff prescription,
  with $\trunc=10,20$.  Dashed lines show the $n=2$ cutoff
  prescription, again with $\trunc=10,20$. For smoothly truncated
  models, ratios of masses outside the virial radius to the virial
  mass, $[M_{\rm tot}-M(<10)]/M(<10)$, are 17\% for ($n$,
  $\trunc$)=(1, 10), 4.6\% for (2, 10), 36\% for (1, 20), and 17\% for
  (2,20). Note that this ratio is infinity for the original NFW halo
  because it has divergent total mass.} 
\label{fig:density}
\end{figure}

With these desiderata in mind, we suggest the following functional form for a
smoothly truncated universal 3-d mass density profile:
\begin{equation}
\rho(r)=\frac{4\deltac \rhoc}
             { \left(\frac{r}{\rs}\right)
               \left(1+\frac{r}{\rs}\right)^2
               \left(1+\left(\frac{r}{\rt}\right)^2\right)^n}=
\frac{M_0}{4\pi r(r+\rs)^2}\,\left(\frac{\rt^2}{r^2+\rt^2}\right)^n.
\label{eq:newdensity}
\end{equation}
Here, $\rt$ is a new parameter which should correspond to the tidal
radius for tidally truncated halos \cite{BT87}. The parameter $n$
controls the sharpness of the truncation; we will investigate the
cases $n=1,2$. For relatively isolated haloes, we expect the tidal radius to be
much larger than the scale radius.  We define $\trunc=\rt/\rs$, expecting
$\trunc>>1$.  Note that $\trunc$ is not necessarily the ``concentration''
parameter, defined as the ratio of a ``virial'' radius to $\rs$.  In the
left-hand panel of Figure~\ref{fig:density}, we plot this profile in the usual
way (with logarithmic axes), and compare with the original NFW profile. We show
the effect of the tidal radius in providing a smooth edge to the halo.  In the
right-hand panel of Figure~\ref{fig:density} we show the integrated projected
mass profile, for the same set of density profiles. In this panel radius is
projected radius, and the logarithmic divergence of the original NFW profile
can be clearly seen.  Projected mass within some appropriate radius is
(approximately) the quantity that is best constrained by gravitational lensing
-- in Section~\ref{sect:obs} we show the predicted observables of gravitational
lensing in more detail.

After some experimentation we found that the form recommended above is indeed
the simplest one that gives an analytic potential while ensuring a
non-diverging total mass for all values of the tidal radius. In the case where
the tidal radius~$\rt$ is outside the scale radius~$\rs$, the $n=1$ profile
falls off as $r^{-5}$, steep enough to mimic a sharp cutoff. If the tidal
radius were to lie inside the scale radius, then the density would decrease as
$r^{-3}$ initially before turning over to $r^{-5}$ outside~$\rs$. Since this
would imply some memory of the original halo after the presumably violent act
of tidal stripping, we suggest that if $\rt<\rs$, the $n=2$ version of the
density profile be used.  For $\rt < \rs$, this profile turns over to $r^{-5}$
at $\rt$, which is effectively a sharp cutoff.  The further turnover to
$r^{-7}$ at $\rs > \rt$ has little effect.

The close agreement of the unstripped halo with the original NFW profile is
comforting.  For example, for a halo with a concentration of 10, if we set
$\rt$ to twice the virial radius $(\trunc=20)$, the masses contained within the
virial radius are the same to within $6\%$.  Of course the total mass of the
truncated halo is 50\% larger than the virial mass, while the total mass of
the untruncated halo (formally) diverges.  We will thus take $\rt=2\rvir$ as
our fiducial tidal radius for an unstripped halo. We note that this
choice of the truncation radius is simply a working assumption in this
paper, and the more appropriate value should eventually be obtained in
$N$-body simulations and observations. 


\section{Lensing by stellar mass in galaxies}
\label{sect:sersic}

The Sersic profile, found to fit well the optical surface brightness~$I$ of
undisturbed galaxies \cite{Ser68}, is
\begin{equation}
I(r)= I_{\rm e} \exp\left[\kappa_n \left(1 - \frac{r}{r_{\rm e}}\right)^{1/n}
\right],
\label{eq:sersic}
\end{equation}
where the effective radius $r_{\rm e}$ is the radius within which half
the flux is contained, and $n$ is the Sersic index. For elliptical
galaxies, an index of around 4 is often seen \cite{deV48}, while the
characteristic exponential profile of galaxy disks corresponds to a
Sersic index of~1. In fact, a broad range of Sersic index values have
been seen in fits to observed galaxy light profiles \cite{Bla05}.
In addition, there has been some arguments that density profiles of CDM haloes
can also be fitted well by the Sersic profile with an index of $2-3$.
\cite{Mer05,T+G05}.

Assuming that stellar mass follows light, we can substitute surface mass
density $\Sigma$ for surface brightness in equation~\ref{eq:sersic}. In the
appendix, we show that the lens potential sourced by this mass distribution
\cite{Car04} is analytically tractable for integer and half-integer $n$.


\section{Predicted observables}
\label{sect:obs}

As we show in the appendix, both density profiles introduced above (truncated
NFW and Sersic) have analytic lens potentials. (The NFW profile has an analytic
three-dimensional potential, which can itself be projected analytically.) The
expressions for the lensing potential, while somewhat lengthy, are rapidly
calculated and differentiable to all orders. In this section we plot some of
these derivatives, pointing out their application in gravitational lens data
modelling.  We note that in the limit of radii beyond the tidal radius the
lensing properties of our model haloes do indeed approach those of a point mass,
as required.


\subsection{Weak lensing}

\begin{figure}
\begin{minipage}[t]{0.48\linewidth}
\centering\epsfig{file=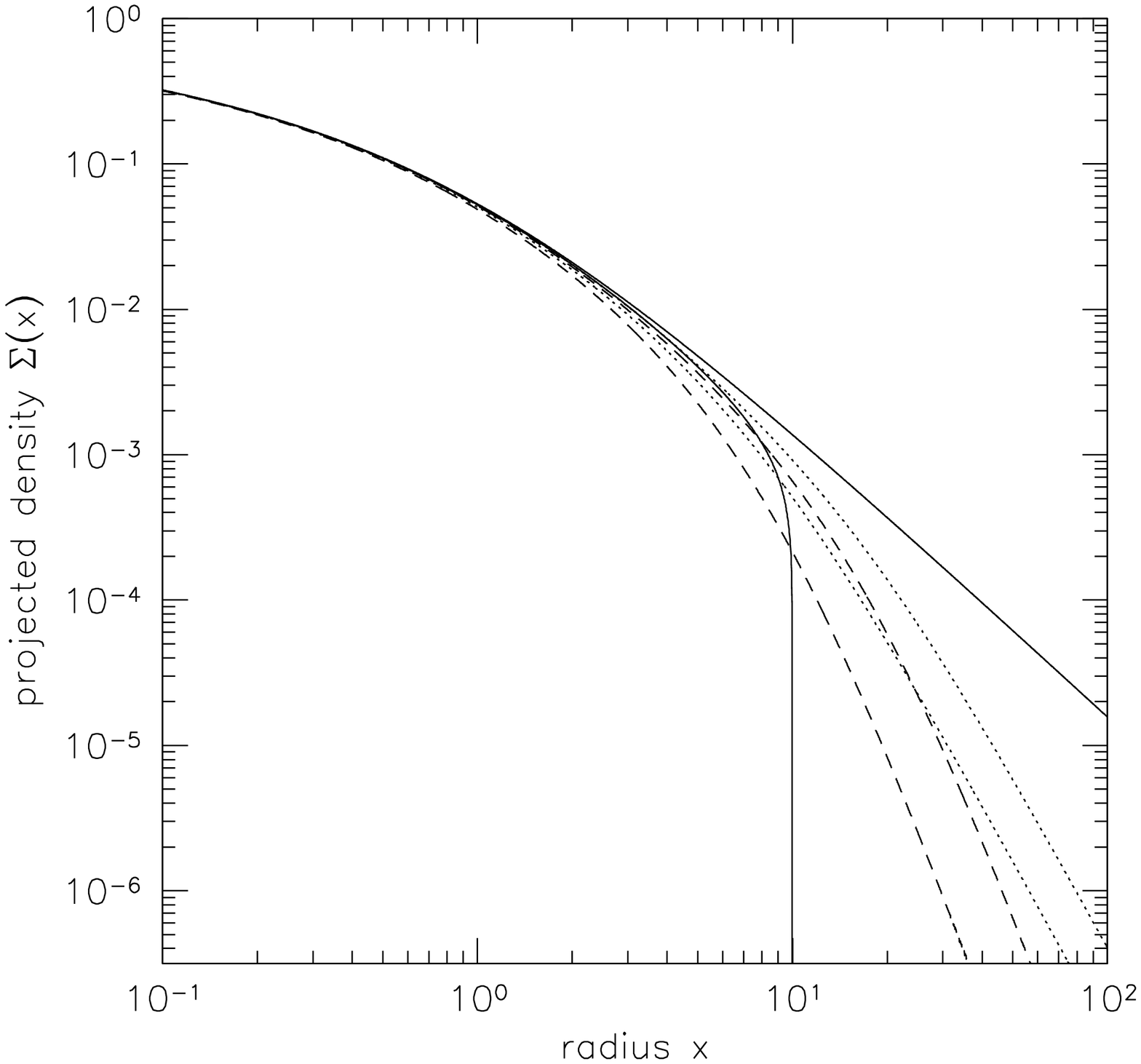,width=0.95\textwidth}
\end{minipage}\hfill
\begin{minipage}[t]{0.48\linewidth}
\centering\epsfig{file=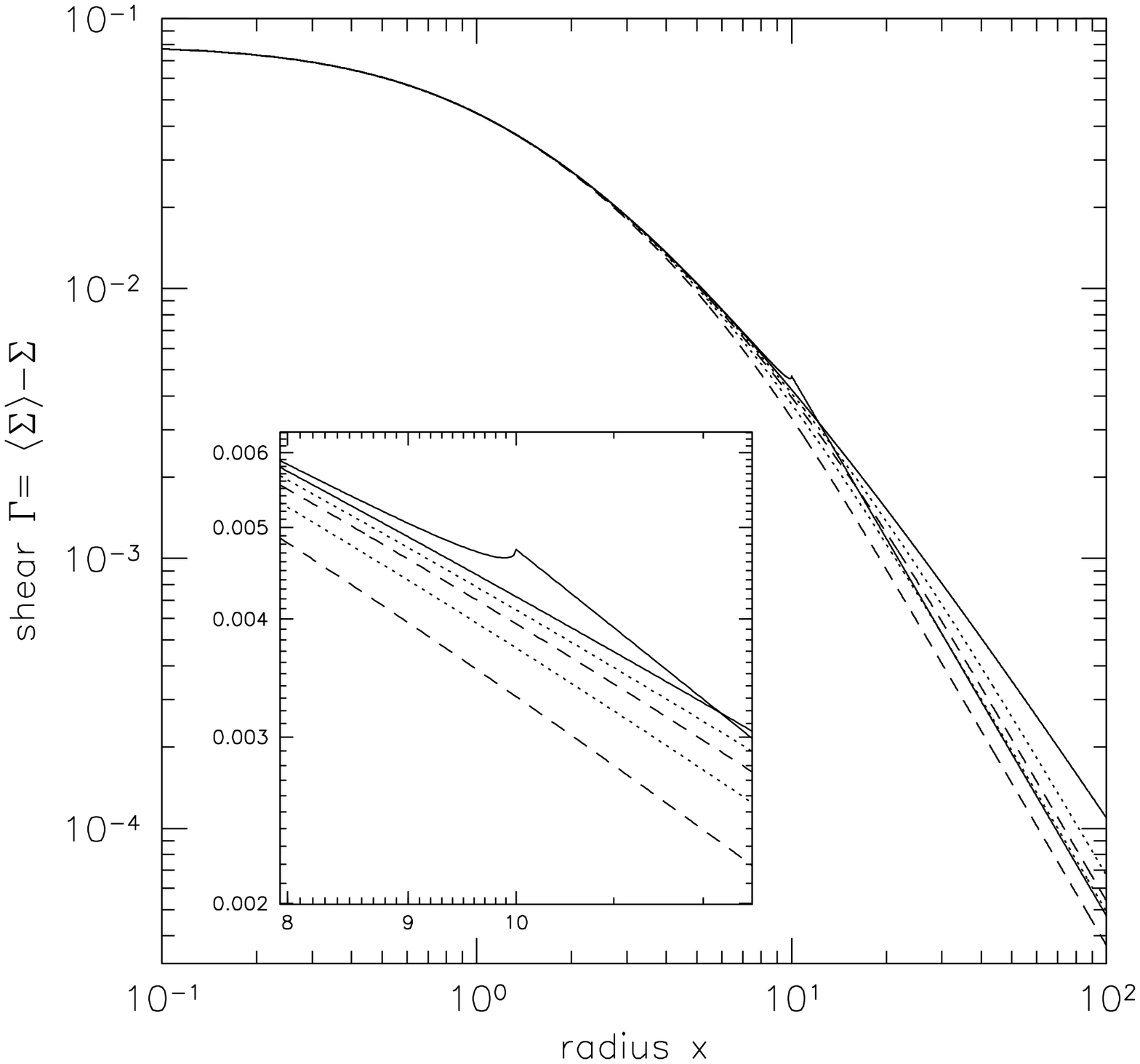,width=0.95\textwidth}
\end{minipage}
\caption{Convergence (left) and shear (right) for NFW haloes.  The curve
  line styles are the same as in Figure~\ref{fig:density}.  Projected mass
  density $\Sigma$ (directly proportional to convergence $\kappa$) is plotted
  on the left, showing that a hard cutoff in density results in a softer, but
  still non-differentiable, cutoff in the convergence.  Actually plotted on the
  right is $\langle\Sigma\rangle-\Sigma$, which is directly proportional to the
  shear $\gamma$ for axisymmetric haloes.  Notice that a hard cutoff in density
  means the shear is finite, but not differentiable at the cutoff radius.}
\label{fig:shear}
\end{figure}

We first address the issue of not truncating the NFW profile on the weak
lensing shear and convergence (see \cite{GL/Sch06} for a good
introduction to these quantities). The lefthand panel of
Figure~\ref{fig:shear} shows the convergence (projected mass density)
profile for the set of haloes first introduced in Figure~\ref{fig:density}.

We see that using a truncated profile with the same virial mass somewhat
reduces the predicted lensing effect. The corresponding shear profiles are
plotted in the right-hand panel of Figure~\ref{fig:shear}.  Taking the central
density profiles to be equal (mimicking a well-constrained
central strong lensing region, for example), for $\trunc=20$ we note
that the virial mass
(mass within 10$\rs$) is 6\% less for the $n=1$ truncated halo.  The shear for
the two profiles only differs by 3\%, however. The total {\em projected} mass
within the projected virial radius is some 12\% lower than that of 
the untruncated profile.  Lastly, the surface
density of the truncated halo is 30\% lower at 10$\rs$.  

The difference in reduced shear $\gamma/|1-\kappa|$ thus depends on
the absolute value of the convergence, relative to the critical
surface density, but can be significant. Very roughly, from
Figure~\ref{fig:shear} we expect that different truncations examined
in this paper can yield $\sim 10\%$ difference in $\gamma$ at around
the virial radius. Although this is smaller than the accuracy of shear
measurements for most massive clusters of galaxies (e.g., \cite{Bro08}),
the accuracy can be reachable in the weak lensing analysis of 
stacked cluster samples (e.g., \cite{Joh07}).

\subsection{Strong and intermediate lensing}

\begin{figure}
\begin{minipage}[t]{0.48\linewidth}
\epsfig{file=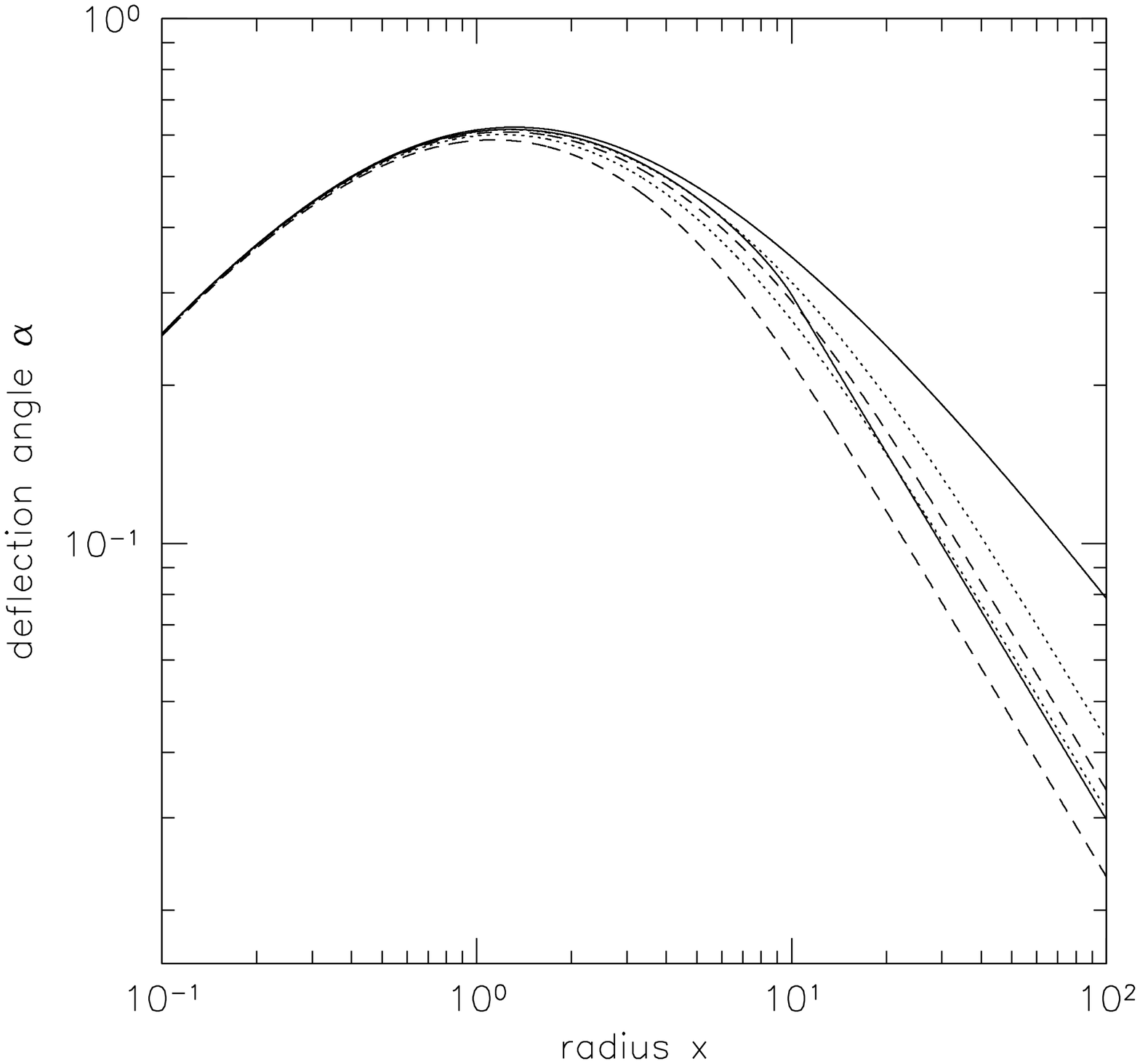,width=0.95\textwidth}
\end{minipage}\hfill
\begin{minipage}[t]{0.48\linewidth}
\epsfig{file=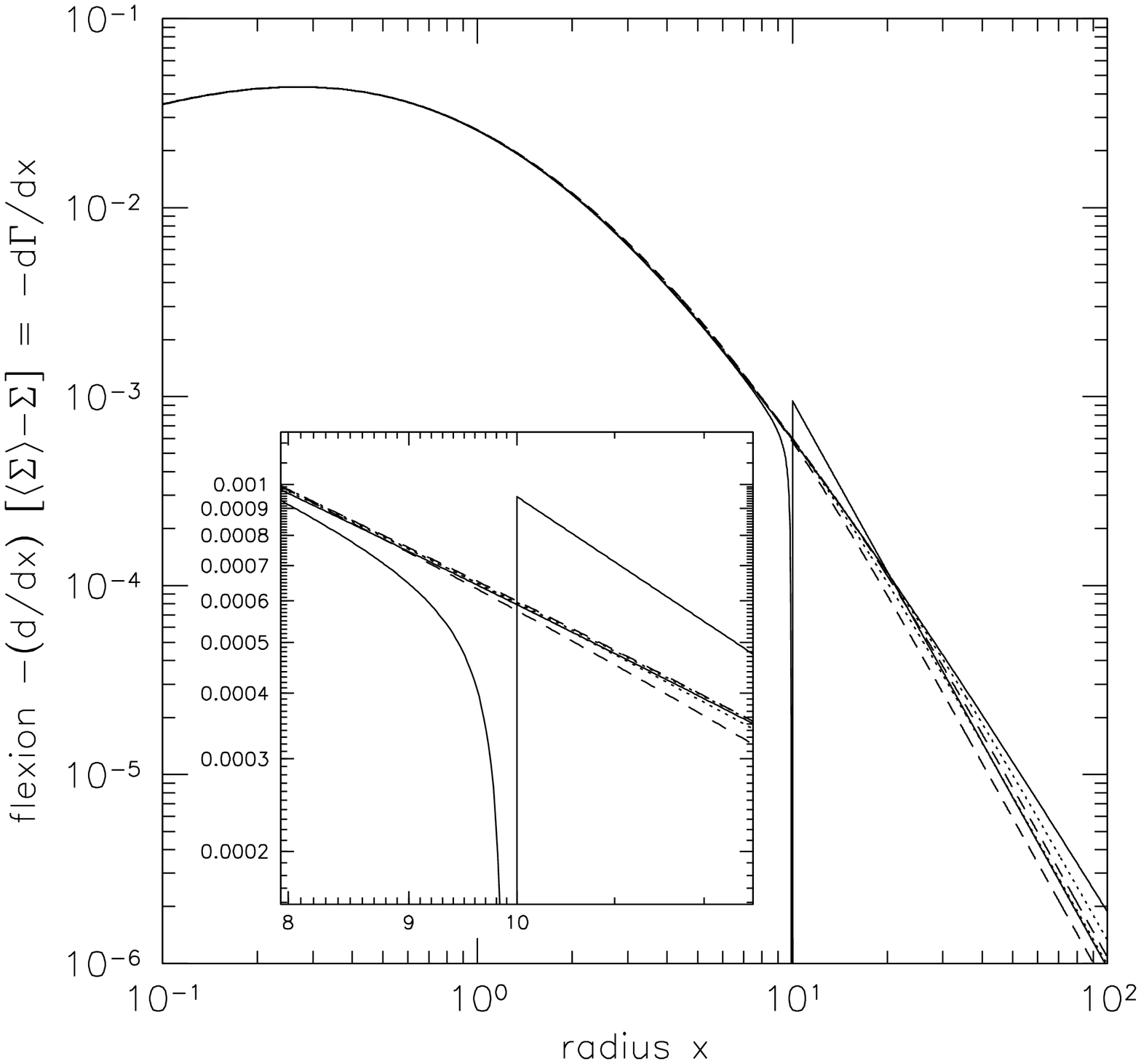,width=0.95\textwidth}
\end{minipage}
\caption{Deflection angle (left) and flexion (right) for NFW haloes. 
Curves are the same as in
Figure~\ref{fig:density}. Notice that the hard cutoff in density causes the
flexion to diverge at the cutoff radius.  As the flexion approaches
$-\infty$, so it changes sign as well.  This can be seen in
Figure~\ref{fig:shear}, where the shear actually starts to increase as the
cutoff radius is approached.}
\label{fig:strong}
\end{figure}

Figure~\ref{fig:strong} shows the amplitude of the deflection angle for
a strongly lensed source. Here we see that using a truncated profile has
very little effect on the deflection angle in the regime where it is measurable
as a multiple-image separation ($r < \rs$). This 
is unsurprising given that the strong lensing is
dominated by the central part of the profile which is, by design,
little changed in our new model.

In contrast, the so-called ``flexion'' may be more strongly affected
by the truncation because it is essentially the higher-order
derivative of the lens potential. In addition it is measurable over a
wide range of scales from the Einstein radius to (in a statistical
sense) the virial radius. In Figure~\ref{fig:strong} we plot the third
derivative of the lens potential: the most interesting component of
flexion for circular symmetry is in fact just this, the radial
gradient of the shear. Marked differences in signal strength arise
when the haloes are truncated. 

\begin{figure}
\begin{minipage}[t]{0.48\linewidth}
\centering\epsfig{file=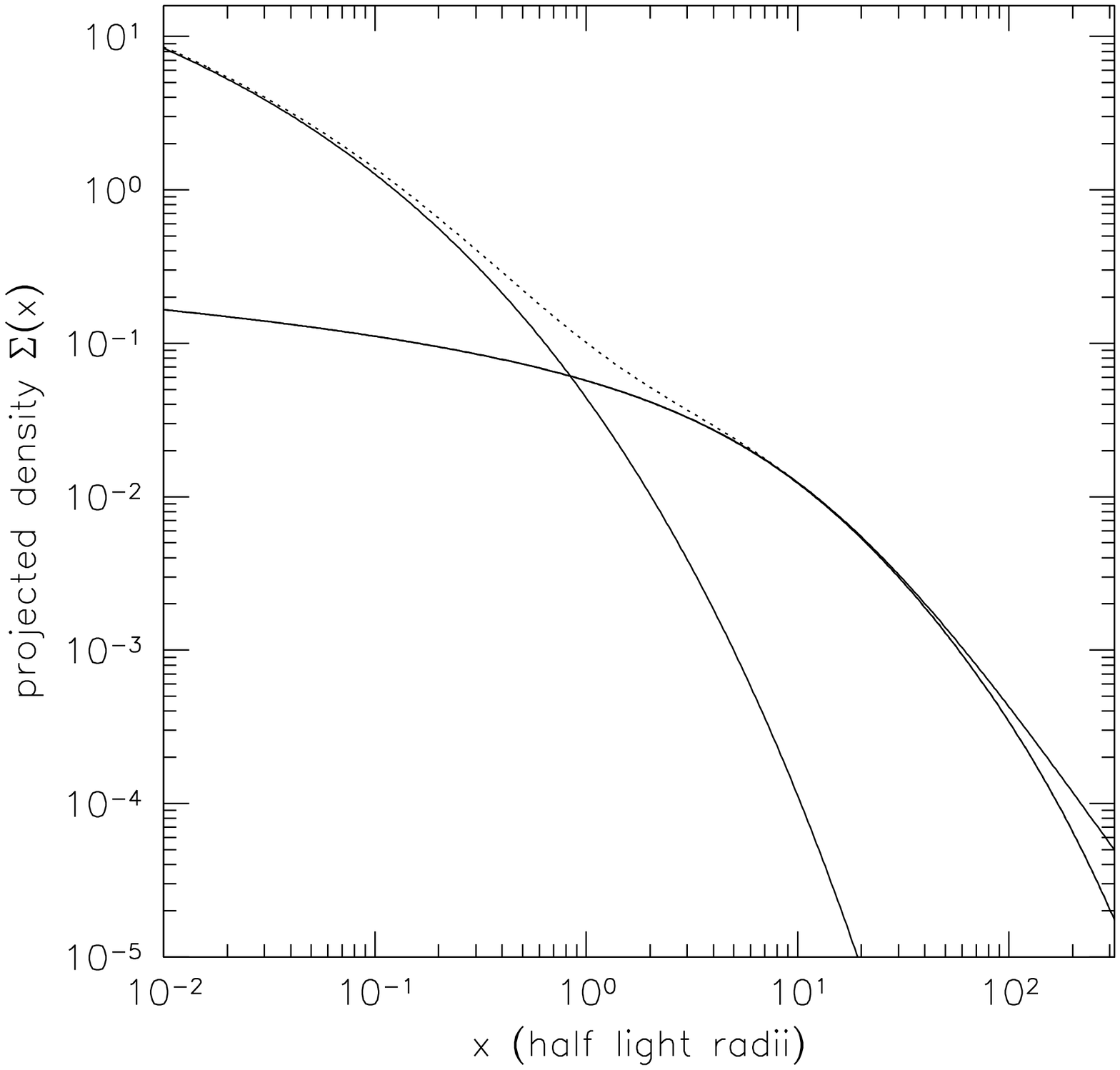,width=0.95\textwidth}
\end{minipage}\hfill
\begin{minipage}[t]{0.48\linewidth}
\centering\epsfig{file=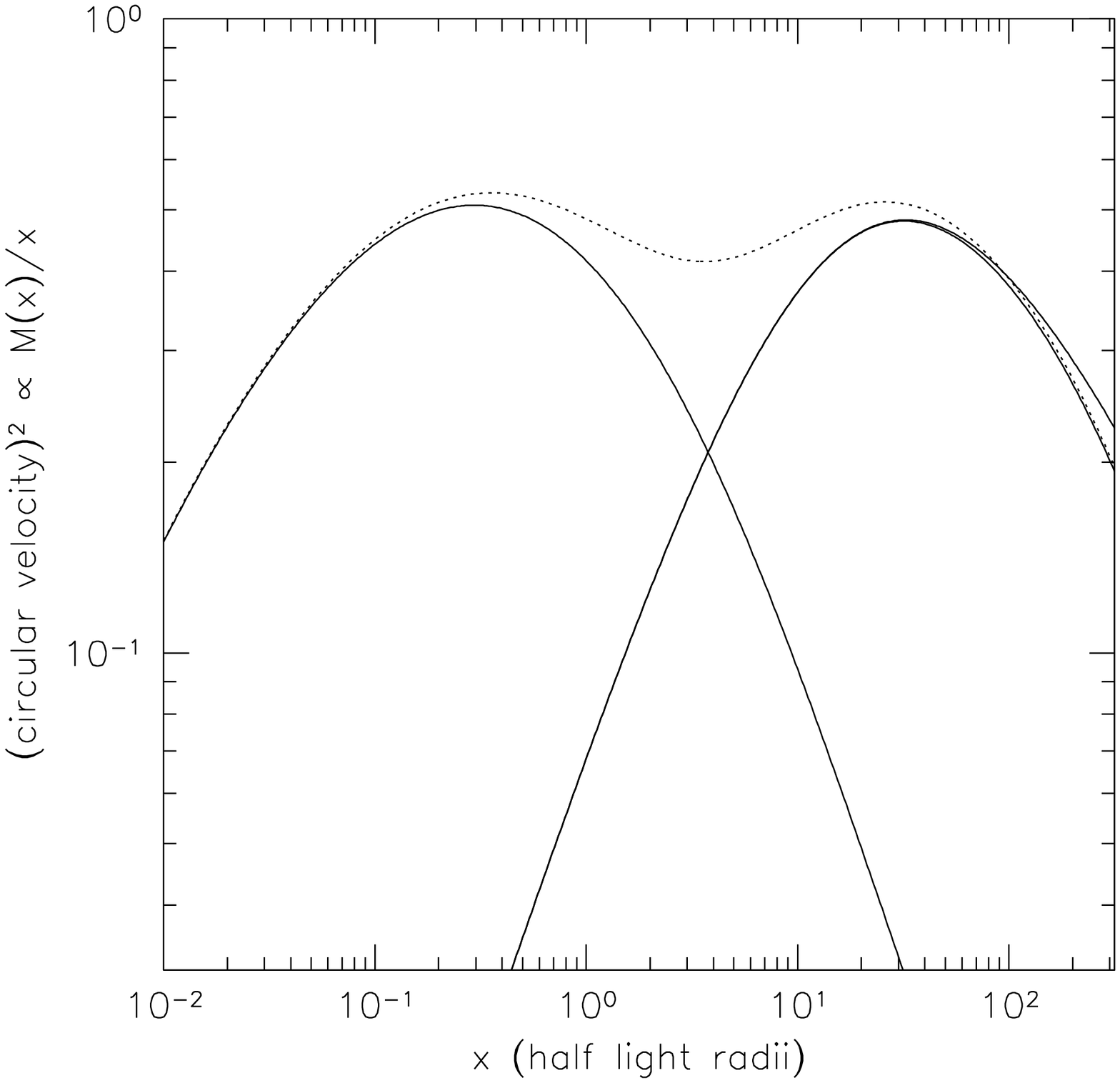,width=0.95\textwidth}
\end{minipage}
\caption{Sersic (n=4) profile combined with NFW profile.  We plot the
  convergence (left) and circular velocity squared $M(x)/x$ (right)
  associated with each component (there are  two curves for the NFW
  profile: one untruncated and one with $\trunc=20$), 
  along with the total (dashed line).  The truncated NFW profile is 70 times
  more massive than the Sersic profile (the virial mass is 50 times larger than
  the stellar mass), and the scale radius is 15 times larger than the half
  light radius.  With these reasonable parameters, it is clear that the total
  profile is nearly isothermal (logarithmic slope of -1 in convergence) around
  the half light radius.  In fact, we find that (for $\trunc=20$), the relation
  $M_{\rm NFW}/M_{\rm deV}=4.75 \,r_s/r_e$ gives a flat region in velocity
  dispersion.}
\label{fig:sersic}
\end{figure}

A plausible model for an elliptical galaxy lens consists of two parts: the
stellar component and the dark halo.  Modelling the stellar component as an
$n=4$ Sersic profile and the dark halo as a truncated NFW profile with
concentration 10 and $\trunc=20$, a reasonable fit to lensing data can
be made \cite{Gav07}.  The salient feature is that the total mass profile is
approximately isothermal.  This can be arranged with the following
prescription (for $\trunc=20$ and $n=4$): 
\begin{equation}
\frac{M_{\rm NFW}}{M_{\rm deV}}\approx 4.75\,\frac{r_s}{r_e}.
\end{equation}
The NFW mass is the total mass.  The virial mass (mass within 10 scale radii in
this case) is 0.66 of the total.  The broad isothermal region obtained by this
prescription is illustrated in Figure~\ref{fig:sersic}. We note that
adiabatic contraction \cite{Zel++80,Blu++86,Gne++04} modifies the NFW
profile, leading to more centrally concentrated profile of dark matter.
However, the effect of adiabatic contraction is most pronounced at the
very center of the halo where the baryonic (Sersic) component is
dominated, and thus its effect on the total mass profile is not
substantial. 

\begin{figure}
\begin{minipage}[t]{0.48\linewidth}
\epsfig{file=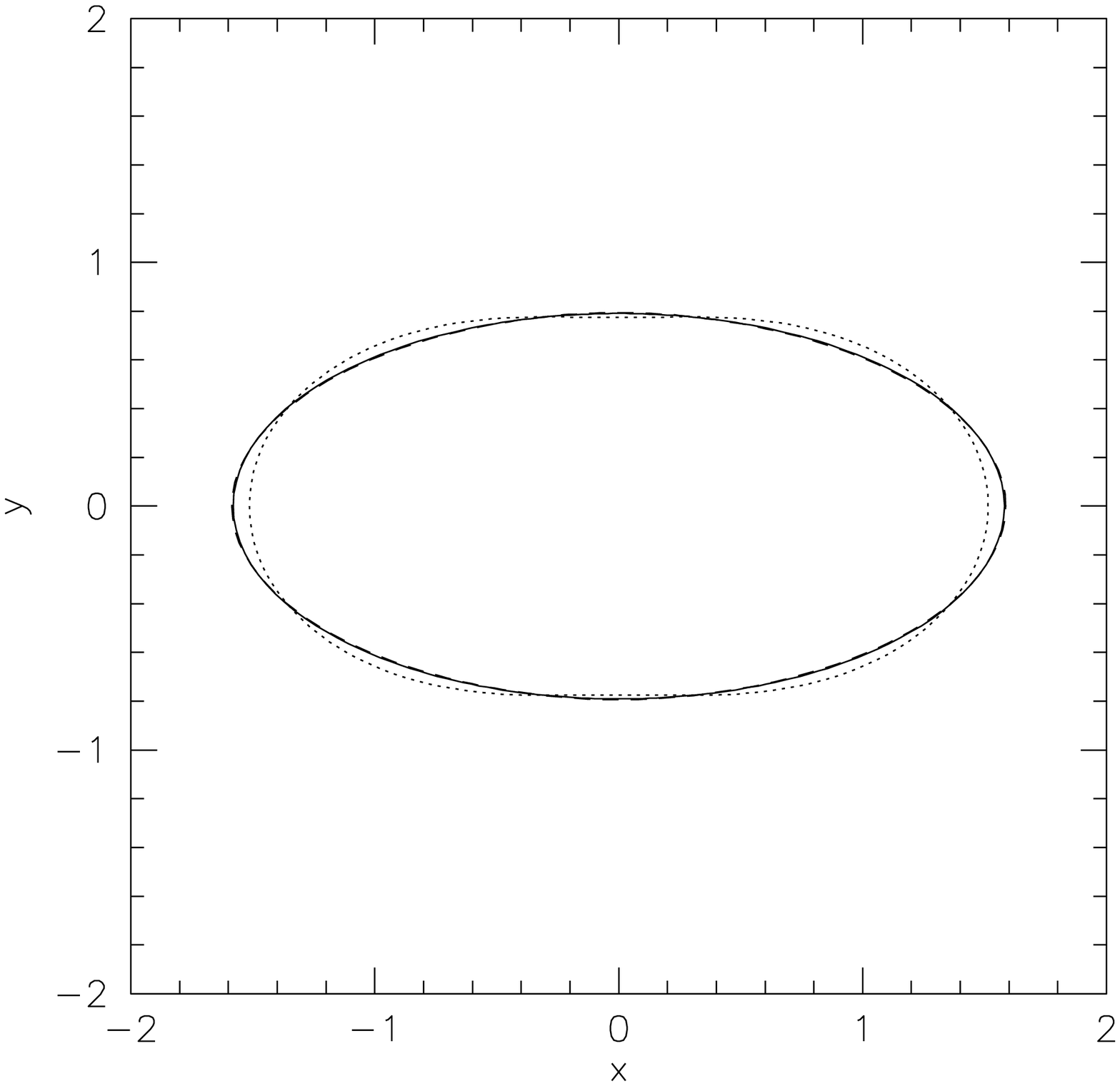,width=0.95\textwidth}
\end{minipage}\hfill
\begin{minipage}[t]{0.48\linewidth}
\epsfig{file=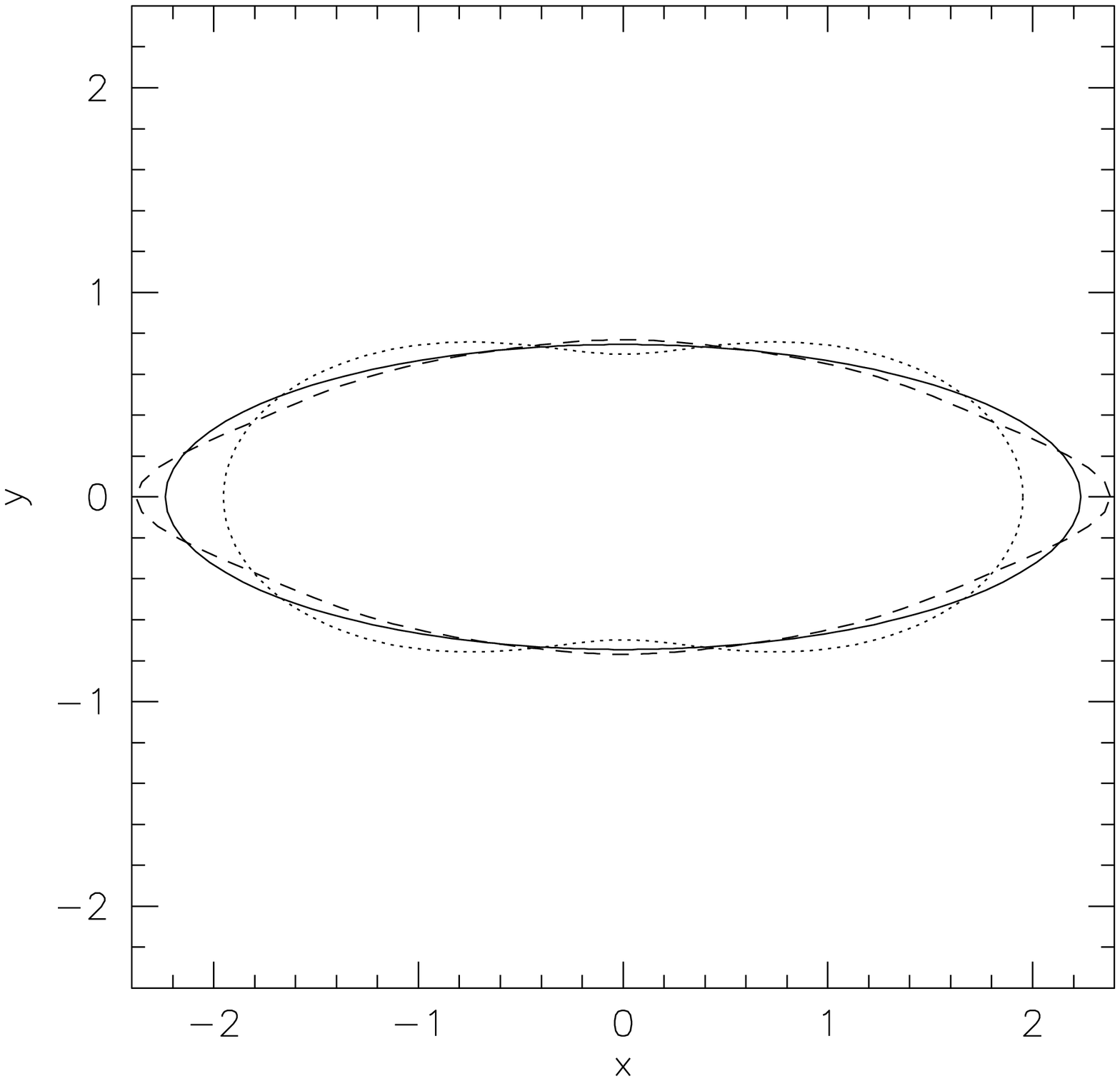,width=0.95\textwidth}
\end{minipage}
\caption{Convergence (isodensity) contours for three stacked elliptical
  potentials.  True ellipses are shown as solid curves, three stacked
  potentials are shown as dashed curves, and single elliptical potentials are
  shown as dotted curves.  The left panel illustrates the case $\epsilon_{\rm
    iso}=0.6$ (axis ratio 2:1), while the right panel illustrates
  $\epsilon_{\rm iso}=0.8$ (axis ratio 3:1).  The slight diskiness of the
  isodensity contours can be seen especially for the $\epsilon_{\rm iso}=0.8$
  case.  The fitting procedure is described in Appendix C.}
\label{fig:ell}
\end{figure}

Finally we move to two dimensions and illustrate the construction of
elliptically symmetric isophotes in the convergence distribution. It has been
noted \cite{K+K93} that the isodensity contours of an
elliptical lens potential can become dumbbell-shaped at low values of the axis
ratio. In the appendix we show how isodensity contours that are elliptical to
third order in the axis ratio can be constructed following a simple recipe. In
Figure~\ref{fig:ell} we illustrate this procedure, showing the constituent
concentric elliptical potentials and the resulting convergence contours.  It is
found that our new model not only avoids the unphysical concave isodensity but
also gives much better fit to the elliptical isodensity than the elliptical
lens potential. For large axis ratios (3:1 is as large as is feasible by our
technique), the isodensity contours are slightly disky.  We note that this
recipe preserves the radial profile of the (self-similar) constituent
potentials. Although the procedure is derived assuming a pure
power-law, the nearly isothermal distribution of the composite model
illustrated in Figure~\ref{fig:sersic} suggests that our prescription
is useful for such composite model, at least as long as the
ellipticities of Sersic and NFW components are similar.

\section{Discussion}
\label{sect:discuss}

We have introduced a simple smoothly-truncated extension of the NFW density
profile. To date the majority of cluster and galaxy lens modelling that has been
performed using the NFW profile has used the untruncated profile. We find that,
if haloes are indeed tidally-moulded leading to the kinds of smooth truncation
that we propose, then the masses of the haloes may have been overestimated by
some 10\% or so during a weak lensing analysis. 
This number pertains to the situation where the inner profile is inferred to be
the same for each model profile, as might be the case when good strong lensing
data are available. We find that the smooth truncation of a halo does not
significantly affect the deflection angles at the image positions (which lie
typically well within the scale radius). If strong lensing data are not
available then the degeneracy between the truncation radius and the halo mass
will give rise to a broader inferred marginal probability distribution for the
halo mass, with the mean shifting to lower values than for the untruncated
profile.

The truncation of galaxy haloes in field galaxy-galaxy lensing is always likely
to be masked by the effects of large scale structure on the outer parts of the
mass profile (the ``two-halo'' term). However, in clusters a 10\% systematic
error is comparable to that introduced in other parts of a current weak (plus
strong) lensing analysis. The uncertain background galaxy redshift distribution,
additional mass along the line of sight, cluster member galaxy contamination,
projection effects, and shear calibration errors can easily be of order 10\% in
the halo mass. However, as survey sizes increase, and the goals of
cosmological cluster-counting experiments become loftier, uncertainties such as
that introduced by halo truncation may become important.

Sharp truncation \cite{T+J03} introduces discontinuities in the shear
that are unlikely to cause problems in data modelling; however, the same may
not be true about flexion, where a singularity appears at the truncation
radius. Smooth truncation (or indeed, no truncation) avoids this problem.

Galaxy-galaxy weak lensing studies within clusters of galaxies have already
succeeded in producing a measurement of a truncation radius \cite{NKS02}. The
combined Sersic plus NFW model currently popular in field galaxy lens modelling
could profitably be applied in a cluster galaxy-galaxy lensing. The photometry
provides extra constraints on the stellar mass part of the density profile
\cite{N+K97}, allowing the dark matter structure of galaxies in clusters to be
probed. The model suggested here would straightforwardly allow strong and
intermediate lensing effects to be incorporated; we are not far from possessing
a useful sample of strong gravitational lenses lying in clusters.

The question of how best to model tidal-truncation of dark matter haloes has been
approached here in a phenomenological and pragmatic way: we wanted an analytic
form for the lens potential. We believe the forms presented here would provide 
good fits to the haloes seen in numerical simulations, based on the
successes of others with very similar profiles \cite{T+B04,D+W05}. 
We have shown that plausible models of gravitational lenses can be constructed
from the superposition of simple analytic profiles, including the generation of
elliptically-symmetric isodensity contours. An interesting extension of this
would be to attempt to build up still more complex mass distributions, from
misaligned and offset building blocks \cite{Mar06}. Again, whether the
haloes and sub-haloes  observed in numerical simulations can be well-enough
approximated by such a model remains to be seen. At present it seems that the
signal-to-noise in Einstein rings is sufficiently high to constrain such a more
complex model \cite{Koo05,Koo++06}.

Finally, we discuss the importance of truncation, and a smooth one at that, in
when simulating lensing effects in large surveys. Gravitational lensing, weak,
intermediate and strong, may be expected to be an important component of
multiple-pronged dark energy investigation.  Simulations of large fields will
play a vital role in improving our understanding of the astrophysical
systematic 
errors present  in cosmic shear, and cluster mass function, measurements. Halo
models are a cheap and efficient way of doing this, capturing the pertinent
physical effects without the need for further CPU-expensive N-body simulations.
However, ray-tracing through halo models does present some technical challenges
\cite{T+J03,O06}. The smooth analytic truncation proposed here allows the mass
budget to be balanced, while allowing  all gravitational lensing effects to be
calculated rapidly to machine precision. In fact, it has been shown
that the shear angular correlation function becomes $\sim 20$\%
smaller if the truncation at around the virial radius is included, and
that the calculation with the truncation shows better agreement with
$N$-body simulations \cite{T+J03}.   

A by-product of the future large optical and radio imaging surveys
will be an interesting sample of strong lenses showing higher-order
catastrophes beyond the usual cusps and folds. These systems provide
very high magnifications, and are very sensitive to the mass structure
in the lens, and as such promise to be interesting
laboratories. However, modelling them will require an accurate
multi-scale approach; we leave the development of this project to
further work.  

The code used in this work is plain ISO C99 and can be freely downloaded from

\vspace{\baselineskip}
\begin{center}
\tt http://kipac.stanford.edu/collab/research/lensing/ample/
\end{center}


\section*{Acknowledgments}

We thank James Taylor, Peter Schneider, Stelios Kazantzidis, and
Masahiro Takada for useful discussions and encouragement.  We also
thank an anonymous referee for many suggestions. This work was
supported in part by the U.S. Department of Energy under contract number
DE-AC02-76SF00515. PJM acknowledges support from the TABASGO foundation
in the form of a research fellowship. 


\bibliographystyle{unsrt}
\bibliography{references}


\appendix

\section{Truncated NFW Profile}

The Navarro, Frenk and White (NFW) profile is given by
\begin{equation}
\rho(r)=\frac{M_0}{4\pi}\frac{1}{r(r+r_s)^2}.
\end{equation}
Defining $x=r/r_s$,
\begin{equation}
\rho(x)=\frac{M_0}{4\pi r_s^3}\frac{1}{x(1+x)^2}.
\end{equation}
This profile describes dark matter haloes well, out to the virial radius.
However it suffers from the deficiency that it has infinite mass.  The
truncation radius is defined to be a factor of $\trunc$ larger than the scale
radius.  With the sole motivation of allowing ``simple'' analytic forms, we
propose the following truncated NFW profile:
\begin{equation}
\rho_T(x)=\frac{M_0}{4\pi r_s^3}\frac{1}{x(1+x)^2}\frac{\trunc^2}{\trunc^2+x^2}.
\end{equation}
This form is quite similar to the NFW profile for $x<\trunc$, i.e.\ inside the
truncation radius.  Furthermore, the total mass is finite,
\begin{equation}
M=M_0\,\frac{\trunc^2}{(\trunc^2+1)^2}
\left[(\trunc^2-1)\ln \trunc+\trunc\pi-(\trunc^2+1)\right].
\end{equation}
In the limit $\trunc\rightarrow\infty$, we recover the logarithmically
divergent mass of the NFW profile.

For purposes of gravitational lensing, we are interested in the projected mass
density.  We first define the function
\begin{equation}
F(x)=\frac{\cos^{-1}(1/x)}{\sqrt{x^2-1}}.
\end{equation}
This function is well defined everywhere: taking the appropriate limit
$F(1)=1$, and for $x<1$, where both the numerator and denominator are purely
imaginary, we choose the branch where $F(x)>0$.  Note that {\tt ArcCos} from
Mathematica and {\tt cacos} from C99 disagree on the sign of $\cos^{-1}(x)$
when $x>1$. Note that in the $x\rightarrow 0$ limit it reduces to
$F(x)=\ln(2/x)$. We also define the following logarithm, which will
appear many times, \begin{equation}
L(x)=\ln\left(\frac{x}{\sqrt{\trunc^2+x^2}+\trunc}\right).
\end{equation}
With these definitions, the projected surface mass density is given by
\begin{eqnarray}
\Sigma(x)&=&r_s\int_{-\infty}^\infty d\ell\;
\rho_T\left(\sqrt{\ell^2+x^2}\right)\\
&=&\frac{M_0}{r_s^2}\,\frac{\trunc^2}{2\pi(\trunc^2+1)^2}
\Bigg\{\frac{\trunc^2+1}{x^2-1}\left[1-F(x)\right]+2 F(x)\nonumber\\
&&-\frac{\pi}{\sqrt{\trunc^2+x^2}}
+\frac{\trunc^2-1}{\trunc\sqrt{\trunc^2+x^2}}\,L(x)\Bigg\}.
\end{eqnarray}
Notice that for $x\rightarrow 1$, the first term requires that a limit be
taken.  As before, in the $\trunc\rightarrow\infty$ limit we recover the result
for the NFW profile,
\begin{equation}
\Sigma(x)=\frac{M_0}{r_s^2}\,\frac{1-F(x)}{2\pi(x^2-1)}
+O\left(\frac{\ln \trunc}{\trunc^2}\right).
\end{equation}
We can derive the convergence $\kappa$ simply by taking
$\kappa=\Sigma/\Sigma_{\rm crit}$.  We will also need the total projected mass
inside radius $x$,
\begin{eqnarray}
M_{\rm proj}(x)&=&r_s^2\int_0^xdx'\,2\pi x'\Sigma(x')\\
&=&M_0\,\frac{\trunc^2}{(\trunc^2+1)^2}
\Bigg\{\left[\trunc^2+1+2(x^2-1)\right]F(x)+
\trunc\pi\nonumber\\
&&+(\trunc^2-1)\ln \trunc
+\sqrt{\trunc^2+x^2}\left[-\pi+
\frac{\trunc^2-1}{\trunc}\,L(x)\right]\Bigg\}.
\end{eqnarray}
We again recover the NFW result in the $\trunc\rightarrow\infty$ limit,
\begin{equation}
M_{\rm proj}(x)=M_0\left(F(x)
+\ln\frac{x}{2}\right)+O\left(\frac{\ln \trunc}{\trunc^2}\right).
\end{equation}
A crucial quantity is the shear of the gravitational field.  The simplest way
to derive it for a circularly symmetric lens is to use the mean projected
surface density inside radius $x$, which is simply $\bar{\Sigma}=M_{\rm
  proj}/\pi r^2$.  Defining $\Gamma=\bar\Sigma-\Sigma$, the shear is then
$\gamma=\Gamma/\Sigma_{\rm crit}$.

We can now derive the lensing potential.  Including the famous factor of two,
and defining $u=x^2$,
\begin{equation}
\psi(u)=\frac{4G}{c^2}\int_0^{\sqrt{u}}\frac{dx'}{x'}\,M_{\rm proj}(x')=
\frac{2G}{c^2}\int_0^u\frac{du'}{u'}\,M_{\rm proj}(\sqrt{u'}).
\end{equation}
The potential is thus
\begin{eqnarray}
\psi(u)&=&\frac{2GM_0}{c^2}\frac{1}{(\trunc^2+1)^2}\Bigg\{2\trunc^2\pi
\left[\trunc-\sqrt{\trunc^2+u}+\trunc\,
\ln\left(\trunc+\sqrt{\trunc^2+u}\right)\right]\nonumber\\
&&+2(\trunc^2-1)\trunc\sqrt{\trunc^2+u}\,L(\sqrt{u})+
\trunc^2(\trunc^2-1)\,L^2(\sqrt{u})\nonumber\\
&&+4\trunc^2(u-1)F(\sqrt{u})+\trunc^2(\trunc^2-1)\left(\cos^{-1}
\frac{1}{\sqrt{u}}\right)^2\nonumber\\
&&+\trunc^2\left[(\trunc^2-1)\ln \trunc-\trunc^2-1\right]\ln u\nonumber\\
&&-\trunc^2\left[(\trunc^2-1)\ln \trunc\,\ln(4\trunc)+2\ln(\trunc/2)-
2\trunc(\trunc-\pi)\ln(2\trunc)\right]\Bigg\}.
\end{eqnarray}
The resulting potential has the correct behavior in the
$\trunc\rightarrow\infty$ limit for all values of $u$,
\begin{equation}
\psi(u)=\frac{2GM_0}{c^2}\left[\left(\cos^{-1}\frac{1}{\sqrt{u}}\right)^2+
\ln^2\left(\frac{\sqrt{u}}{2}\right)\right]+
O\left(\frac{\ln \trunc}{\trunc^2}\right).
\end{equation}
We also verify that for $u>>\trunc$, the potential is that of a point mass of
the correct total mass (up to an irrelevant constant $A$),
\begin{eqnarray}
\psi(u)&=&A+\frac{4GM_0}{c^2}\,\frac{\trunc^2}{(\trunc^2+1)^2}
\left[(\trunc^2-1)\ln \trunc+\trunc\pi-
(\trunc^2+1)\right]\ln\sqrt{u}\nonumber\\
&&+O\left(\frac{\trunc^2}{u}\right)\\
&=&A+\frac{4GM}{c^2}\,\ln\sqrt{u}+O\left(\frac{\trunc^2}{u}\right).
\end{eqnarray}

We can now calculate the derivatives of $\psi$.  The first is obvious,
\begin{equation}
\psi'(u)=\frac{2G}{c^2}\,\frac{M_{\rm proj}(\sqrt{u})}{u}.
\end{equation}
Next is slightly messier,
\begin{eqnarray}
\psi''(u)&=&\frac{2G}{c^2}\left[-\frac{M_{\rm proj}(\sqrt{u})}{u^2}+
\frac{\pi r_s^2\Sigma(\sqrt{u})}{u}\right]
=-\frac{2\pi Gr_s^2\Gamma}{c^2u}\\
&=&\frac{2GM_0}{c^2}\,\frac{\trunc^2}{2(\trunc^2+1)^2u^2}
\Bigg[2(1-u-\trunc^2)\,F(\sqrt{u})
-2(\trunc^2-1)\ln \trunc\nonumber\\
&&-\frac{(\trunc^2-1)(u+2\trunc^2)}{\trunc\sqrt{\trunc^2+u}}\,L(\sqrt{u})
+\frac{\pi(\sqrt{\trunc^2+u}-\trunc)^2}{\sqrt{\trunc^2+u}}\nonumber\\
&&{+}\frac{(\trunc^2+1)u}{u-1}\left[1-F(\sqrt{u})\right]\Bigg].
\end{eqnarray}
Again, the NFW result appears in the infinite $\trunc$ limit,
\begin{equation}
\psi''(u)=\frac{2GM_0}{c^2}\left[\frac{u+(2-3u)F(\sqrt{u})}{2u^2(u-1)}
-\frac{1}{u^2}\ln\left(\frac{\sqrt{u}}{2}\right)+O(\trunc^{-2})\right].
\end{equation}
At this point, we can calculate the time delay, deflection angle, convergence,
shear, and magnification of the truncated NFW lens.  We will calculate one more
derivative, deriving the so-called ``flexion'' (e.g., \cite{gol05,gol06}).
\begin{eqnarray}
\psi'''(u)&=&\frac{2G}{c^2}\left[\frac{2M_{\rm proj}(\sqrt{u})}{u^3}-
\frac{2\pi r_s^2\Sigma(\sqrt{u})}{u^2}+
\frac{\pi r_s^2\Sigma'(\sqrt{u})}{2u^{3/2}}\right]\\
&=&-\frac{2\pi Gr_s^2}{c^2u^2}\left(u\Gamma'-\Gamma\right)\\
&=&\frac{2GM_0}{c^2}\,\frac{\trunc^2}{2(\trunc^2+1)^2u^3}\Bigg\{
\frac{3(\trunc^2+1)^2u}{2(\trunc^2+u)(u-1)^2}\left[uF(\sqrt{u})-1\right]\nonumber\\
&&+\frac{u}{2(\trunc^2+u)(u-1)}\left[(\trunc^2+1-2(u-1))uF(\sqrt{u})\right.\nonumber\\
&&\left.-(\trunc^2+1)((u-1)+2\trunc^2+3)\right]\nonumber\\
&&+\frac{2(\trunc^2+1)}{u-1}\left[F(\sqrt{u})-u\right]
+2\left[2(u-1)+3\trunc^2+1\right]F(\sqrt{u})\nonumber\\
&&+\frac{\trunc^2-1}{2\trunc(\trunc^2+u)^{3/2}}
\left[3(3\trunc^2+u)(\trunc^2+u)-\trunc^4\right]
L(\sqrt{u})\nonumber\\
&&+4(\trunc^2-1)\ln \trunc
-\frac{2\pi(\sqrt{\trunc^2+u}-\trunc)^2}{\sqrt{\trunc^2+u}}
+\frac{\pi u^2}{2(\trunc^2+u)^{3/2}}
\Bigg\}.
\end{eqnarray}
Expanding about $\trunc=\infty$, we find the NFW result,
\begin{eqnarray}
\psi'''(u)&=&\frac{2GM_0}{c^2}\left[\frac{(15u^2-20u+8)F(\sqrt{u})+3u-6u^2}
{4u^3(u-1)^2}\right.\nonumber\\
&&\left.+2\ln\left(\frac{\sqrt{u}}{2}\right)+O(\trunc^{-2})\right].
\end{eqnarray}

For comparison, we institute a sharper cutoff,
\begin{equation}
\rho_T(x)=\frac{M_0}{4\pi r_s^3}\frac{1}{x(1+x)^2}\frac{\trunc^4}
{(\trunc^2+x^2)^2}.
\end{equation}
The total mass of this profile is
\begin{equation}
M=M_0\frac{\trunc^2}{2(\trunc^2+1)^3}\left[2\trunc^2(\trunc^2-3)
\ln \trunc-(3\trunc^2-1)(\trunc^2+1-\trunc\pi)\right].
\end{equation}
The projected mass density is
\begin{eqnarray}
\Sigma(x)&=&\frac{M_0}{r_s^2}\,\frac{\trunc^4}{4\pi(\trunc^2+1)^3}
\Bigg\{\frac{2(\trunc^2+1)}{x^2-1}\left[1-F(x)\right]+8F(x)
\nonumber\\
&&+\frac{\trunc^4-1}{\trunc^2(\trunc^2+x^2)}-\frac{\pi[4(\trunc^2+x^2)+\trunc^2+1]}
{(\trunc^2+x^2)^{3/2}}\nonumber\\
&&+\frac{\trunc^2(\trunc^4-1)+(\trunc^2+x^2)(3\trunc^4-6\trunc^2-1)}
{\trunc^3(\trunc^2+x^2)^{3/2}}\,L(x)\Bigg\}.
\end{eqnarray}
The total projected mass is obtained as before,
\begin{eqnarray}
M_{\rm proj}(x)&=&
M_0\,\frac{\trunc^4}{2(\trunc^2+1)^3}\Bigg\{2[\trunc^2+1+4(x^2-1)]F(x)
\nonumber\\
&&+\frac{1}{\trunc}
\left[\pi(3\trunc^2-1)+2\trunc(\trunc^2-3)\ln \trunc\right]\nonumber\\
&&+\frac{1}{\trunc^3\sqrt{\trunc^2+x^2}}\left[-\trunc^3\pi
[4(\trunc^2+x^2)-\trunc^2-1]\right.\nonumber\\
&&\left.+\left[-\trunc^2(\trunc^4-1)+(\trunc^2+x^2)(3\trunc^4-6\trunc^2-1)\right]
L(x)\right]\Bigg\}.
\end{eqnarray}
Finally, the lensing potential is
\begin{eqnarray}
\psi(u)&=&\frac{2GM_0}{c^2}\frac{1}{2(\trunc^2+1)^3}\Bigg\{2\trunc^3\pi
\left[(3\trunc^2-1)\ln\left(\trunc+\sqrt{\trunc^2+u}\right)\right.\nonumber\\
&&\left.-4\trunc\sqrt{\trunc^2+u}\right]+2(3\trunc^4-6\trunc^2-1)\trunc\sqrt{\trunc^2+u}\,L(\sqrt{u})\nonumber\\
&&+2\trunc^4(\trunc^2-3)\,L^2(\sqrt{u})+16\trunc^4(u-1)F(\sqrt{u})\nonumber\\
&&+2\trunc^4(\trunc^2-3)\left(\cos^{-1}
\frac{1}{\sqrt{u}}\right)^2\nonumber\\
&&+\trunc^2\left[2\trunc^2(\trunc^2-3)\ln \trunc-3\trunc^4-2\trunc^2+1\right]
\ln u\nonumber\\
&&+2\trunc^2\Big[\trunc^2(4\trunc\pi+(\trunc^2-3)\ln^22+8\ln 2)\nonumber\\
&&-\ln(2\trunc)(1+6\trunc^2-3\trunc^4+\trunc^2(\trunc^2-3)\ln(2\trunc))\nonumber\\
&&-\trunc\pi(3\trunc^2-1)\ln(2\trunc)\Big]\Bigg\}.
\end{eqnarray}
As before,
\begin{equation}
\psi'(u)=\frac{2G}{c^2}\,\frac{M_{\rm proj}(\sqrt{u})}{u}.
\end{equation}
The second derivative follows,
\begin{eqnarray}
\psi''(u)&=&\frac{2GM_0}{c^2}\,\frac{\trunc^4}{4(\trunc^2+1)^3u^2}
\Bigg\{\left[-8(u-1)+2(1-3\trunc^2)\right]\,F(\sqrt{u})\nonumber\\
&&-4(\trunc^2-3)\ln \trunc+\frac{2(\trunc^2+1)}{u-1}\left[1-F(\sqrt{u})\right]+8\nonumber\\
&&+\frac{(\trunc^2+u)(3\trunc^4-6\trunc^2-1)-
\trunc^2(\trunc^4-1)}{\trunc^2(\trunc^2+u)}\nonumber\\
&&+\frac{2u(3\trunc^4-6\trunc^2-1)}
{\trunc^3(\trunc^2+u)^{1/2}}\,L(\sqrt{u})\nonumber\\
&&-\frac{(3u+2\trunc^2)[(\trunc^2+u^2)(3\trunc^4-6\trunc^2-1)-\trunc^2(\trunc^4-1)]}
{\trunc^3(\trunc^2+u)^{3/2}}\,L(\sqrt{u})\nonumber\\
&&-\frac{2\pi}{\trunc}(3\trunc^2-1)\nonumber\\
&&+\frac{\pi}{(\trunc^2+u)^{3/2}}
\left[2(\trunc^2+u)(3\trunc^2-1)+u(4(\trunc^2+u)-\trunc^2-1)\right]
\Bigg\}.
\end{eqnarray}
Finally, the third derivative,
\begin{eqnarray}
\psi'''(u)&=&\frac{2GM_0}{c^2}\,\frac{\trunc^4}{8(\trunc^2+1)^3u^3}\Bigg\{
\Bigg[\frac{6(\trunc^2+1)}{(u-1)^2}+\frac{20\trunc^2+12}
{u-1}+30\trunc^2\nonumber\\
&&-18+24(u-1)\Bigg]\,F(\sqrt{u})-\frac{1}{(\trunc^2+u)^2}\Bigg[\frac{6(\trunc^2+1)^3}{(u-1)^2}\nonumber\\
&&+\frac{2(\trunc^2+1)^2(5\trunc^2+7)}{u-1}+4(\trunc^2+1)^2(u+\trunc^2+2)\nonumber\\
&&+(2+3\trunc^2)u^2\Bigg]-\frac{8(\trunc^2+1)}{u-1}-32
+16(\trunc^2-3)\ln \trunc\nonumber\\
&&+\frac{u^2}{\trunc^2(\trunc^2+u)^2}+
\frac{8\pi}{\trunc}(3\trunc^2-1)\nonumber\\
&&-\frac{4\left[(\trunc^2+u)(3\trunc^4-6\trunc^2-1)-
\trunc^2(\trunc^4-1)\right]}{\trunc^2(\trunc^2+u)}\nonumber\\
&&-\frac{\pi\left[24\trunc^6+3u^2(4u-5)+
5\trunc^2u(9u-4)+\trunc^4(60u-8)\right]}{(\trunc^2+u)^{5/2}}\nonumber\\
&&+\frac{16\trunc^{10}+30\trunc^6u(u-4)+9\trunc^4u^2(u-10)}
{\trunc^3(\trunc^2+u)^{5/2}}\,L(\sqrt{u})\nonumber\\
&&+\frac{-3(6\trunc^2+1)u^3+8\trunc^8(5u-6)}
{\trunc^3(\trunc^2+u)^{5/2}}\,L(\sqrt{u})
\Bigg\}
\end{eqnarray}
All of these results reduce to the pure NFW case in the limit
$\trunc\rightarrow\infty$.

As a final comparison, we reproduce the \cite{T+J03} formulas for a
sharp cutoff at $x=\trunc$.  In addition, we will derive the third derivative
of the potential.  First, we define the auxiliary function
\begin{equation}
T(x)=\frac{1}{\sqrt{x^2-1}}\left(\tan^{-1}\frac{\sqrt{\trunc^2-x^2}}
{\sqrt{x^2-1}}
-\tan^{-1}\frac{\sqrt{\trunc^2-x^2}}{\trunc\sqrt{x^2-1}}\right),
\end{equation}
choosing the branch where $T(x)$ is positive.  This function is well defined
everywhere provided we take the limit at $x=1$, finding
\begin{equation}
T(1)=\sqrt{\frac{\trunc-1}{\trunc+1}}
\end{equation}
Here, {\tt ArcTan} from Mathematica and {\tt catan} from C99 agree on the
branch of $\tan^{-1}(x)$ for purely imaginary argument.
\begin{eqnarray}
\Sigma(x)&=&\frac{M_0}{r_s^2}\frac{1}{2\pi}
\left[\frac{1}{x^2-1}\left(
\frac{\sqrt{\trunc^2-x^2}}{\trunc+1}-T(x)\right)\right].
\end{eqnarray}
\begin{eqnarray}
M_{\rm proj}(x)&=&M_0
\Bigg[\ln(1+\trunc)-\frac{\trunc}{1+\trunc}+\frac{\sqrt{\trunc^2-x^2}}{\trunc+1}
\nonumber\\
&&-\tanh^{-1}\frac{\sqrt{\trunc^2-x^2}}{\trunc}+T(x)\Bigg].
\end{eqnarray}
Note that the total mass is
\begin{equation}
M=M_{\rm proj}(\trunc)=M_0\left[\ln(1+\trunc)-\frac{\trunc}{1+\trunc}\right].
\end{equation}
Note that these are valid only for $x\leq \trunc$. For $x>\trunc$, we
simply have $\Sigma(x)=0$ and $M_{\rm proj}(x)=M$.

We find that the potential involves at least the polylogarithm
$\Li_2$; we do not find a simple expression.  We start with the first
derivative, 
\begin{equation}
\psi'(u)=\frac{2G}{c^2}\,\frac{M_{\rm proj}(\sqrt{u})}{u}.
\end{equation}
Assuming $u<\trunc^2$,
\begin{eqnarray}
\psi''(u)&=&-\frac{2GM_0}{c^2}\frac{1}{u^2}
\Bigg\{\ln(1+\trunc)-\frac{\trunc}{1+\trunc}+\frac{\sqrt{\trunc^2-u}}{\trunc+1}
\nonumber\\
&&-\tanh^{-1}\frac{\sqrt{\trunc^2-u}}{\trunc}
-\frac{u\sqrt{\trunc^2-u}}{2(\trunc+1)(u-1)}+\frac{3u-2}{2(u-1)}\,T(\sqrt{u})
\Bigg\},
\end{eqnarray}
\begin{eqnarray}
\psi'''(u)&=&\frac{2GM_0}{c^2}\frac{2}{u^3}
\Bigg\{\ln(1+\trunc)-\frac{\trunc}{1+\trunc}+\frac{\sqrt{\trunc^2-u}}{\trunc+1}
\nonumber\\
&&-\tanh^{-1}\frac{\sqrt{\trunc^2-u}}{\trunc}
-\frac{u\sqrt{\trunc^2-u}}{2(\trunc+1)(u-1)}\nonumber\\
&&+\frac{15u^2-20u+8}{8(u-1)^2}\,T(\sqrt{u})\nonumber\\
&&+\frac{u[(u-1)\trunc+u(2+u)-\trunc^2(1+2u)]}{8(1+\trunc)
\sqrt{\trunc^2-u}(u-1)^2}\Bigg\}.
\end{eqnarray}
Note that as $u\rightarrow \trunc^2$, $\psi'''$ diverges (see also Figure
\ref{fig:strong}). This is clearly because the surface mass density of
this model is not smooth at $u=\trunc^2$. On the other hand, for
$u>\trunc^2$, the derivatives can easily be computed as
$\psi'(u)=2GM/c^2u$, $\psi''(u)=-2GM/c^2u^2$,  and
$\psi'''(u)=4GM/c^2u^3$.


\section{The Sersic Profile}

The Sersic profile describes the light distribution of elliptical galaxies, and
it also can be made to fit the mass distribution of haloes \cite{Mer05}.  This
profile is defined in projection,
\begin{equation}
\Sigma(x)=\frac{M_0}{r_0^2\pi(2n)!}\exp\left(-x^{1/n}\right),
\end{equation}
where $r_0$ is a scale radius and $x=r/r_0$.  The effective radius $r_e$ which
contains half of the projected mass (or light) can be determined numerically.
For the de Vaucouleurs profile with $n=4$, $r_e=3459.5r_0$.  The
three-dimensional density distribution can be obtained by Abel inversion,
\begin{eqnarray}
\rho(x)&=&-\frac{1}{\pi r_0}\int_x^\infty \frac{dx'}{\sqrt{x'^2-x^2}}
\frac{d\Sigma}{dx'}\nonumber\\
&=&-\frac{1}{\pi r_0}\int_0^{\pi/2}du\,\sec u\,
\frac{d\Sigma}{dx'}\left(x'=x\sec u\right),
\end{eqnarray}
and we find in practice that the integral with finite range works well
numerically.  As before, we need the total projected mass inside radius $x$,
\begin{equation}
M_{\rm proj}(x)=M_0\left(1-\frac{\Gamma(2n,x^{1/n})}{\Gamma(2n)}\right).
\end{equation}
The total mass is
\begin{equation}
M=M_0.
\end{equation}
The lensing potential can be expressed as a generalized hypergeometric
function \cite{Car04},
\begin{equation}
\psi(u)=\frac{2GM_0}{c^2}\frac{u}{(2n)!}\;
\hypgeo2F2\left(2n,2n;2n+1,2n+1;-u^{1/2n}\right),
\end{equation}
where again $u=x^2$.  When $2n$ takes integer values, these results can be
expressed in simpler terms as follows,
\begin{eqnarray}
\psi(u)&=&\frac{2GM_0}{c^2}\,2n
\Biggl[\ln u^{1/2n}+\gamma-\expint\left(-u^{1/2n}\right)
-a^n_0\nonumber\\
&&+\exp\left(-u^{1/2n}\right)\sum_{j=0}^{2n-2}a_j^n\,
\frac{u^{j/2n}}{j!}\Biggr],
\end{eqnarray}
with
\begin{equation}
a_j^n=\sum_{k=j+1}^{2n-1}\frac{1}{k},
\end{equation}
and $\expint(x)$ being the exponential integral function.  Now we calculate the
derivatives of $\psi$.  As before,
\begin{equation}
\psi'(u)=\frac{2G}{c^2}\,\frac{M_{\rm proj}(\sqrt{u})}{u}.
\end{equation}
When $2n$ is an integer, we can express this in terms of elementary functions,
\begin{equation}
\psi'(u)=\frac{2GM_0}{c^2}\,\frac{1}{u}\left(1-\exp\left(-u^{1/2n}\right)
\sum_{j=0}^{2n-1}\frac{u^{j/2n}}{j!}\right).
\end{equation}
The second derivative is
\begin{eqnarray}
\psi''(u)&=&\frac{2GM_0}{c^2}\,
\Biggl[-\frac{1}{u^2}\left(1-\exp\left(-u^{1/2n}\right)\sum_{j=0}^{2n-1}
\frac{u^{j/2n}}{j!}\right)\nonumber\\
&&+\frac{1}{u(2n)!}\,\exp\left(-u^{1/2n}\right)\Biggr].
\end{eqnarray}
Finally, the third derivative is
\begin{eqnarray}
\psi'''(u)&=&
\frac{2GM_0}{c^2}\,\Biggl[\frac{2}{u^3}\left(1-\exp\left(-u^{1/2n}\right)
\sum_{j=0}^{2n-1}\frac{u^{j/2n}}{j!}\right)\nonumber\\
&&-\frac{1}{u^2(2n)!}\,
\exp\left(-u^{1/2n}\right)\left(2+\frac{u^{1/2n}}{2n}\right)\Biggr].
\end{eqnarray}


\section{Elliptical Potentials}

The power of considering $\psi$ as a function of $u=x^2$ and not $x$ becomes
apparent when considering elliptical potentials.  We will describe how the
various observable quantities are obtained from the potential, and how these
are modified for the case of an elliptical potential.

The deflection angle is just the gradient of the potential,
\begin{equation}
\vec{\alpha}=\vec{\nabla}\psi(u)=\psi'(u)\vec{\nabla} u.
\end{equation}
The lens equation is given by
\begin{equation}
\vec{s}=\vec{r}-\vec{\alpha}(\vec{r}),
\end{equation}
and thus the magnification matrix is
\begin{equation}
\mathbf J=\mathbf I -\vec{\nabla}\otimes\vec{\nabla}\psi(u).
\end{equation}
For this we need the second derivatives of $\psi$,
\begin{equation}
\vec{\nabla}\otimes\vec{\nabla}\psi(u)=
\psi'(u)\vec{\nabla}\otimes\vec{\nabla} u+
\psi''(u)\vec{\nabla}u \otimes\vec{\nabla} u.
\end{equation}
Now we can define $u$ so that the isopotential lines are ellipses and not
circles:
\begin{eqnarray}
u&=&(1-\epsilon)x^2+(1+\epsilon)y^2,\\
v&=&-(1-\epsilon)x^2+(1+\epsilon)y^2,\\
\vec{\nabla} u &=&
\left(\begin{array}{c}2(1-\epsilon)x\\2(1+\epsilon)y\end{array}\right),\\
\vec{\nabla}\otimes\vec{\nabla} u &=&
\left(\begin{array}{cc}2(1-\epsilon)&0\\0&2(1+\epsilon)\end{array}\right),
\end{eqnarray}
and higher derivatives, such as the third needed for flexion, vanish.
Deviations from ellipticity manifest as dependence on the variable $v$.

It is well known that elliptical isopotentials can lead to unphysical projected
mass densities.  For the singular isothermal sphere, $\epsilon>0.2$ gives
isodensity contours that become peanut-shaped.  We propose a simple resolution
to this problem that will be adequate for most purposes: add several elliptical
potentials at the same location.  We find that three potentials, one with
ellipticity $\epsilon$, one with ellipticity $f\epsilon$, and one circular
potential, can be summed in such a way as to give nearly elliptical isodensity
contours with ellipticities as large as $\epsilon=0.8$, in other words axis
ratios as large as three to one.  Consider a power-law density profile
$\rho(r)\propto r^{-\gamma}$, yielding a potential $\psi(u)\propto
u^{(3-\gamma)/2}$.  We combine three such potentials,
\begin{equation}
\psi(u,v)=a_\epsilon\psi_\epsilon(u)+a_{f\epsilon}\psi_{f\epsilon}(u)+
a_0\psi_0(u),
\end{equation}
where the subscript indicates the ellipticity.  Note that all three have purely
elliptical isopotentials, but ellipses of one ellipticity are expressed with
both the $u$ and $v$ variables of a different ellipticity.  Expanding in
$\epsilon$, the $v$ dependence can be canceled up through order $\epsilon^2$
with the following choice of coefficients,
\begin{eqnarray}
a_{f\epsilon}&=&\frac{2a_\epsilon}{f[(5-\gamma)-f(7-\gamma)]},\\
a_0&=& (1-f)\left(\frac{7-\gamma}{3-\gamma}\,f-1\right)a_{f\epsilon}.
\end{eqnarray}
A fraction $f=1/2$ produces the best results in most cases,
\begin{eqnarray}
a_{\epsilon/2}&=&\frac{8}{3-\gamma}\,a_\epsilon,\\
a_0&=& \frac{1+\gamma}{4(3-\gamma)}\,a_{\epsilon/2}=\frac{2(1+\gamma)}
{(3-\gamma)^2}\,a_\epsilon.
\end{eqnarray}
For potentials more complex than a pure power law, the cancellation is only
exact at a single radius and hence the construction of the elliptical model
which fits the wide range in radii is not trivial. We leave this issue
for future work.  

It is also well known that elliptical isopotentials can lead to an inferred
surface mass density that is negative.  This occurs for any finite ellipticity
when the asymptotic density profile is $r^{-3}$ or steeper.  The total inferred
mass is always positive however, and the negative surface mass densities
inferred are small enough to not be a serious concern.

We make a note of the total mass of elliptical profiles here.  Assume a profile
described by a lensing potential $\psi(u)$ with $u$ defined as above.  In
general, the surface density (inferred from the convergence) is given by
\begin{equation}
\Sigma\propto\psi'+\psi''(u+\epsilon v).
\end{equation}
If this inferred surface density contains a finite total mass, the total mass
of the lens as a function of the ellipticity of the isopotentials is given by
\begin{equation}
M_{\rm tot}(\epsilon)=\frac{1}{\sqrt{1-\epsilon^2}}\,M_{\rm tot}(\epsilon=0).
\end{equation}

More practically, we relate the following projected elliptical power-law mass
density
\begin{equation}
\Sigma=\frac{3-\gamma}{2}\left[(1-\epsilon_{\rm iso})x^2+
(1+\epsilon_{\rm iso})y^2\right]^{(1-\gamma)/2},
\end{equation}
with that derived by combining three elliptical isopotentials
\begin{equation}
\psi=b_\epsilon\left[a_\epsilon\psi_\epsilon+a_{\epsilon/2}\psi_{\epsilon/2}+
a_0\psi_0\right],
\end{equation}
where 
\begin{equation}
\psi_\epsilon=\frac{[(1-\epsilon)x^2+(1+\epsilon)y^2]^{(3-\gamma)/2}}{3-\gamma}.
\end{equation}
We find that the following fitting forms connect these two models in the
range $1.2\lesssim\gamma\lesssim2.9$: 
\begin{equation}
\epsilon = \epsilon_{\rm iso}^{}+[0.05(2.1-\gamma)^2+0.257]
\epsilon_{\rm iso}^{0.4(1.8-\gamma)^2+2.9},
\end{equation}
\begin{equation}
b_\epsilon = 1+0.193(-0.9+\gamma)^{1.38}\epsilon_{\rm iso}^2+
0.0121(0.1+\gamma)^{4.06}\epsilon_{\rm iso}^6.
\end{equation}
This relation breaks down rapidly for $\epsilon_{\rm iso}>0.8$, corresponding
to an axis ratio of 3:1. Two isodensity contours match with $\lesssim0.6\%$ level at
$\epsilon_{\rm iso}<0.6$ and $\lesssim3\%$ level at $\epsilon_{\rm iso}<0.8$.



\end{document}